\documentclass[10pt,twocolumn]{article}
\usepackage[a4paper,top=0.72in,bottom=0.72in,left=0.70in,right=0.70in]{geometry}
\usepackage[T1]{fontenc}
\usepackage[utf8]{inputenc}
\usepackage{newtxtext,newtxmath}
\usepackage{graphicx}
\usepackage{amsmath}
\usepackage{microtype}
\usepackage{caption}
\usepackage{booktabs}
\usepackage{tabularx}
\usepackage{cuted}
\usepackage{stfloats}
\usepackage{titlesec}
\usepackage{xcolor}
\usepackage{eso-pic}
\usepackage[numbers,super,sort&compress]{natbib}
\usepackage[colorlinks=true,linkcolor=blue,citecolor=blue,urlcolor=blue]{hyperref}
\graphicspath{{./}}
\titleformat{\section}{\large\bfseries}{\thesection}{0.5em}{}
\titlespacing*{\section}{0pt}{8pt}{3pt}
\titleformat{\paragraph}[runin]{\bfseries}{ }{0pt}{}[.]

\title{\Large From spin splitting to projected mass in altermagnetic Chern matter}
\author{Gyanti Prakash Moharana\textsuperscript{*}\\\small Institute of Physics, Sachivalaya Marg, Bhubaneswar 751005, India}
\date{}

\begin{document}
\pagestyle{plain}
\AddToShipoutPictureFG*{%
  \AtPageLowerLeft{%
    \hspace{0.70in}\raisebox{0.35in}{\color{blue}\footnotesize\textsuperscript{*}Correspondence: \href{mailto:gyantiprakashm@gmail.com}{gyantiprakashm@gmail.com}}%
  }%
}

\makeatletter
\twocolumn[
\begin{@twocolumnfalse}
\maketitle
\vspace{-1.2em}
\begin{center}
\begin{minipage}{0.96\textwidth}
\small
\textbf{Abstract}\par\vspace{0.35em}
Altermagnetic spin splitting alone does not define Chern matter. The relevant object is the exchange mass projected onto Hall-active surface, valley, orbital or interface sectors. We formulate this projected-mass criterion for compensated magnetic topology. The resulting two-channel $(C,\mathcal{A})$ diagnostic separates hidden compensated Hall responses from additive altermagnetic quantum anomalous Hall phases in a global insulating gap. It also guides interface, thickness and materials design strategies.
\end{minipage}
\end{center}
\vspace{0.4em}
\end{@twocolumnfalse}
]
\makeatother

\setcounter{dbltopnumber}{2}
\renewcommand{\dbltopfraction}{0.92}
\renewcommand{\textfraction}{0.05}
\renewcommand{\floatpagefraction}{0.80}

\begin{figure*}[t]
\centering
\includegraphics[width=0.98\textwidth]{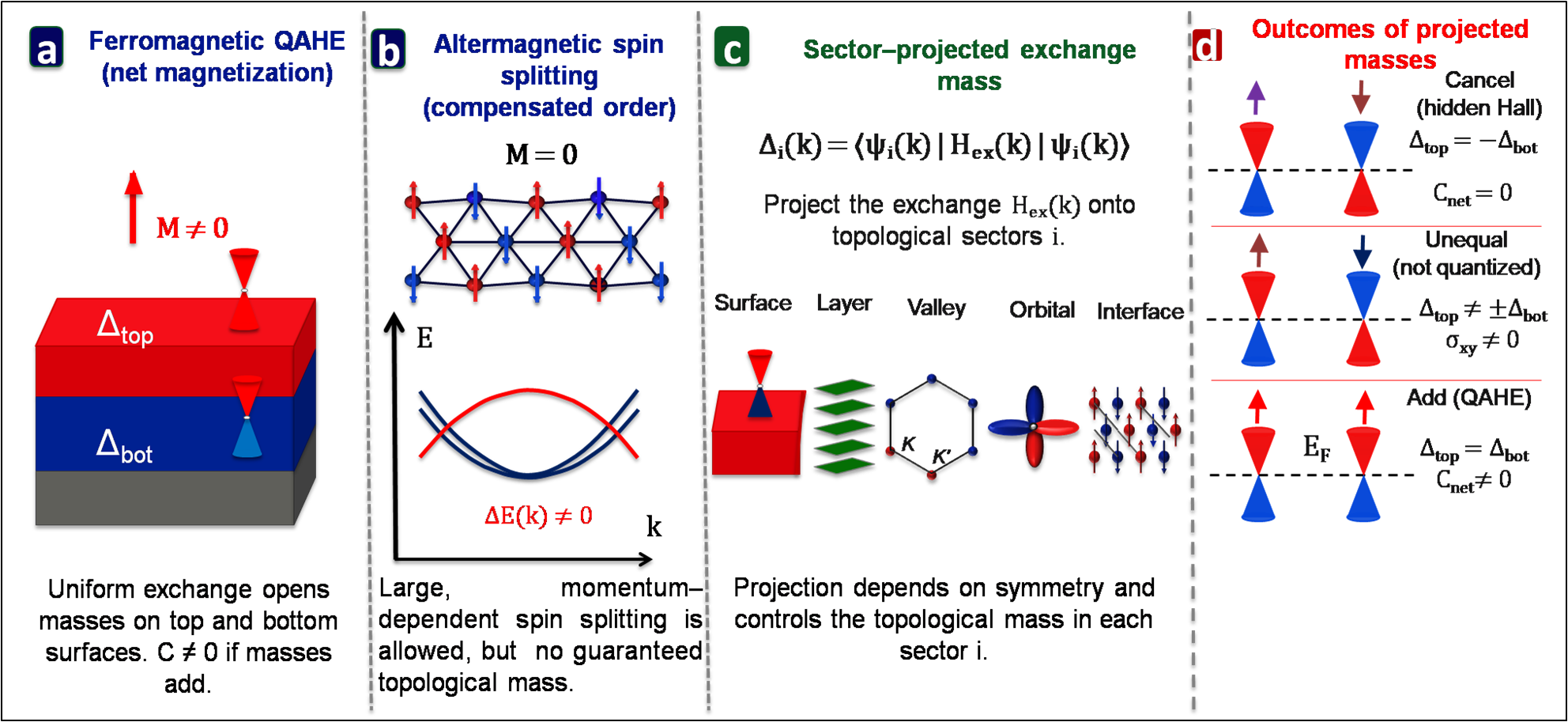}
\caption{\textbf{Exchange projection, not spin splitting alone, generates a Chern mass.}
Magnetic topological phases are controlled by how exchange gaps Hall-active bands. In the altermagnetic route, the net moment may remain compensated, while crystal symmetry structures the exchange field in momentum and sector space. The decisive object is the sector-projected exchange mass, $\Delta_i(\mathbf{k})=\langle\psi_i(\mathbf{k})|H_{\rm ex}(\mathbf{k})|\psi_i(\mathbf{k})\rangle$, with $i$ denoting a surface, layer, valley, orbital or interface sector. Interface normal, registry, strain, capping, gating and thickness decide whether the sectoral masses cancel, become unequal or add. Only the additive case yields a non-zero Chern number in a global insulating gap.}\label{fig:exchange-projection}
\end{figure*}

\section{Introduction}
The quantum anomalous Hall effect (QAHE) is a direct manifestation of exchange-driven topological band reconstruction. Broken time-reversal symmetry $\mathcal T$ and SOC create an insulating Chern gap whose occupied-band Berry curvature $\Omega_n(\mathbf{k})$ integrates to an integer Chern number $C$. A chiral edge channel then conducts at zero applied field, ideally with $\sigma_{xy}=Ce^2/h$ and $R_{xy}=h/(Ce^2)$.\cite{ref1,ref2,ref3,ref4} However, the experimental constraints are stringent. The original Cr-doped (Bi,Sb)$_2$Te$_3$ observation was in the millikelvin regime. High-precision V-doped films reached $\sigma_{xy}=0.9998\pm0.0006\,e^2/h$ at $25\,{\rm mK}$. Five-septuple-layer MnBi$_2$Te$_4$ showed zero-field QAHE at $1.4\,{\rm K}$.\cite{ref1,ref2,ref5} These values define a transport requirement at the Fermi energy $E_{\rm F}$.

A finite exchange splitting alone does not establish a Hall-active mass. Exchange gaps,  surface conduction and magnetic domains decide whether an anomalous Hall signal becomes a quantized CI state.\cite{ref5,ref6,ref7,ref8} Disorder broadening $\Gamma$, residual carriers and chemical-potential control set the transport threshold and the observable QAHE temperature $T_{\rm QAHE}$.\cite{ref9,ref10,ref11} Metrological precision also requires magnetic-noise stability and robust quantization.\cite{ref12,ref13,ref14,ref15}

The emergence of altermagnetism makes this issue even more subtle and experimentally consequential. It combines collinear compensated order with spin-split bands. Within the spin-group framework, the two oppositely polarized magnetic sublattices are related by a spin-reversing crystallographic operation $[R_s \Vert g]$, where $R_s$ denotes spin reversal and $g$ is the corresponding crystal-space operation. They are not connected by $\mathcal{P}\mathcal{T}$ or by translation--time symmetry $\tau\mathcal T$. Crystallographic symmetry fully constrains the momentum-space texture of the spin splitting $\Delta E(\mathbf{k})$. 
Its sign reversals, symmetry-protected nodal planes, and angular harmonics are all dictated by the underlying crystal symmetry.\cite{ref16,ref17,ref18,ref19} Modern symmetry-based classifications identify altermagnetism as a distinct form of magnetic order. Accordingly, altermagnets are fundamentally differentiated from anomalous-Hall antiferromagnets, weak ferromagnets, and conventional $\tau\mathcal T$ antiferromagnets.\cite{ref20,ref21,ref22} Experiments on $\alpha$-MnTe and CrSb have moved altermagnetism into a quantitative band-structure regime. In $\alpha$-MnTe, the reported splitting reaches $370\,{\rm meV}$ and follows the transition near $T_{\rm N}=267\,{\rm K}$. CrSb shows near-$E_{\rm F}$ altermagnetic splitting of order $0.6$--$1.0\,{\rm eV}$.\cite{ref23,ref24,ref25} Spin precession and spin-biased quantum spin Hall response make the altermagnetic splitting transport-active. Thus, it acts as a symmetry-governed transport quantity rather than a passive band feature.\cite{ref26,ref27}

This observation does not directly imply compensated Chern matter. Spin compensation in the spectrum does not necessarily enforce compensation of Berry curvature or Chern number. A Chern insulator does not require macroscopic magnetization, but it does require a time-reversal-breaking mass in the low-energy topological sector. An altermagnet may exhibit strong band splitting while remaining decoupled from the relevant Dirac, valley, or interface states. Conversely, even a weak exchange component can dominate the Berry curvature if it projects onto the appropriate topological mass channel.

Here, ``altermagnetic Chern matter'' denotes a compensated magnetic topological phase in which altermagnetic exchange projects onto Hall-active low-energy bands. The resulting sectoral Chern masses may either compensate or add to a net Chern response. We use ``compensated topology'' for cases in which sector-resolved Berry curvature and local Hall responses exist although the ordinary Hall conductance cancels. The claim is that altermagnets do not automatically generate QAHE. The diagnostic pair $(C,\mathcal{A})$ records the additive Chern response and the hidden compensated Hall order, respectively.

\section{Altermagnetism and Hall topology: separate standards}
Altermagnetism should be identified through magnetic symmetry analysis and momentum-resolved spectroscopy before any topological or transport interpretation is invoked. A credible candidate must show compensated order, a symmetry-consistent N\'eel vector $\mathbf{n}$ and spin splitting tied to that order. Momentum-resolved spectroscopy, neutron diffraction, dichroism, microscopy and transport provide complementary constraints. A single anomalous Hall loop does not establish altermagnetic symmetry. Weak ferromagnetism, non-collinearity, strain-induced symmetry lowering, domains and stoichiometric drift can reproduce partial transport signatures. RuO$_2$ demonstrates the risk: magnetic order, symmetry assignment and sample stoichiometry must be checked independently before a material is used as an altermagnetic reference.\cite{ref28,ref29,ref30}

Topological Hall matter is constrained by an independent bulk--boundary criterion. A metallic anomalous Hall effect is not a QAHE. A Chern insulator needs a global gap, an integer Chern number, suppressed longitudinal conduction and edge transport with the expected chirality. Magnetic topological insulators establish this mass-sign constraint explicitly. The sign of the exchange mass, not merely the presence of magnetic order, determines whether top and bottom surface contributions add or cancel.\cite{ref5,ref6,ref31,ref12}

An altermagnetic Chern claim must pass both tests. The exchange source must be genuinely compensated and altermagnetic. It must also enter the topological mass channel. Many plausible claims fail at this second step. Large spin splitting is not enough unless the exchange operator has the correct symmetry, orientation and wave-function overlap to gap the Hall-active state. Table~\ref{tab:evidence-hierarchy} summarizes the evidence hierarchy for establishing altermagnetic Chern matter.

\begin{table*}[t]
\centering
\caption{\textbf{Evidence hierarchy for altermagnetic Chern matter.} Each level removes a common false-positive interpretation.}
\label{tab:evidence-hierarchy}
\small
\renewcommand{\arraystretch}{1.18}
\begin{tabular}{p{0.22\textwidth}p{0.31\textwidth}p{0.35\textwidth}}
\hline
\textbf{Evidence level} & \textbf{Required observation} & \textbf{What it establishes} \\
\hline
Compensated magnetic order & Magnetic diffraction, magnetic imaging, XMCD/XMLD, $\mu$SR or related probes resolving $\mathbf{n}$, domains and stoichiometry & The response is not from a ferromagnetic impurity phase, a weak moment or an unresolved magnetic texture. \\
Altermagnetic spin splitting & Momentum-resolved spectroscopy or symmetry-resolved transport showing $\Delta E(\mathbf{k})$ with altermagnetic symmetry linked to $\mathbf{n}$ & The compensated order breaks time reversal in the electronic structure. \\
Projected topological mass & SOC-enabled gap opening in the intended surface, valley, layer, orbital or interface sector, with orientation, thickness or symmetry controls & The exchange couples to the Chern-mass channel rather than only splitting metallic bands. \\
Compensated Hall order & Gate-, layer-, valley-, surface- or optical probes revealing sectoral Hall response & Berry curvature is present even when the ordinary Hall channel cancels. \\
Quantized altermagnetic QAHE & Zero-field $R_{xy}=h/(Ce^2)$ plateau, suppressed $R_{xx}$, insulating bulk and chiral edge transport & Compensated exchange has become additive in the Chern channel. \\
\hline
\end{tabular}

\end{table*}

\section{Projected mass criterion}
For an active topological state $|\psi_i(\mathbf{k})\rangle$ in sector $i$, define the Hall-active exchange mass as
\begin{equation}
\Delta_i(\mathbf{k})=\langle\psi_i(\mathbf{k})|H_{\rm ex}(\mathbf{k})|\psi_i(\mathbf{k})\rangle .
\label{eq:sectoral-mass}
\end{equation}
Here the sector may be a surface $s$, layer $\ell$, valley $v$, orbital representation $\alpha$, interface channel or hybridized combination. Equation~\eqref{eq:sectoral-mass} defines the projected exchange mass entering the low-energy Hall sector. It is not meant to replace a multiband calculation. In a low-energy Dirac description, this projection determines the sector mass $m_i$ whose sign controls the local Hall contribution. If the SOC-entangled projected component in the topological mass channel vanishes, altermagnetic spin splitting alone cannot produce a Chern response. The operational requirement is $E_{\rm F}$ inside a mobility gap, $|\Delta_i|$ larger than disorder broadening $\Gamma$, and a mass pattern whose signs survive the symmetry operations left by the interface.

The logic illustrated schematically in Fig.~\ref{fig:exchange-projection} is familiar from topological-insulator surfaces. An exchange component normal to the spin texture can open a mass gap. An in-plane component may only shift the Dirac cone. Altermagnets add a new constraint: crystal symmetry can reverse or reshape the exchange from one sector to its partner. Interface orientation, registry, strain, capping, gating and thickness therefore become topology variables. They are not secondary sample details.

\begin{figure*}[t]
\centering
\includegraphics[width=0.98\textwidth]{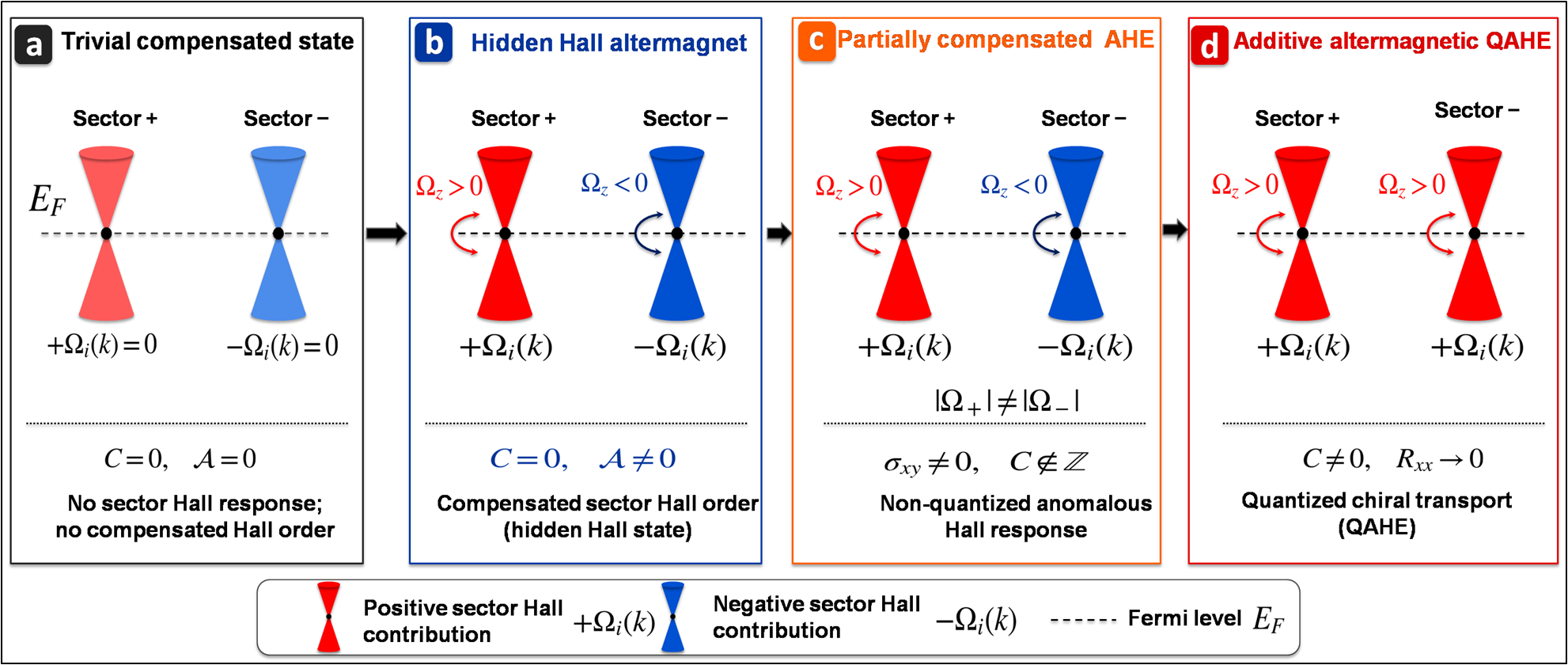}
\caption{\textbf{Two-channel topology separates hidden Hall order from additive QAHE.}
The measured Chern number $C$ shows the additive Hall channel. The diagnostic $\mathcal{A}$ records compensated sectoral Hall order. Signs and weights of the projected masses are modified by gate bias, strain, interface registry, thickness and disorder. Non-zero integer $C$, global insulating gap and chiral edge transport are necessary for the quantized altermagnetic QAHE.}\label{fig:two-channel}
\end{figure*}

\section{Compensated topology and the $(C,\mathcal{A})$ diagnostic}
A single Chern number is sufficient when all Berry curvature contributes to the measured Hall response. It is less transparent when symmetry-paired sectors carry opposite Hall currents. To expose this structure, assign each sector a chirality $\chi_i$. This sign fixes how a positive mass contributes to Hall conductance. Assign also a compensation parity $\eta_i$. This parity fixes how the sector transforms under the operation exchanging altermagnetic partners. The additive channel is
\begin{equation}
C=\frac{1}{2}\sum_i \chi_i\,\operatorname{sgn}(\Delta_i),
\label{eq:C}
\end{equation}
whereas the compensated channel is
\begin{equation}
\mathcal{A}=\frac{1}{2}\sum_i \chi_i\eta_i\,\operatorname{sgn}(\Delta_i).
\label{eq:A}
\end{equation}
On a lattice, $C$ is an integer for a true insulator. By contrast, $\mathcal{A}$ is a sector-projected Hall functional, not a stable bulk invariant. It is well defined only for symmetry-fixed or weakly hybridized sector blocks, and reduces in controlled limits to mirror-Chern, valley-Hall, layer-Hall, spin-Chern or axion surface-mass descriptions. Thus $\mathcal{A}$ tracks compensation-odd Berry curvature, while $C$ tracks additive quantized Hall topology.

Figure~\ref{fig:two-channel} schematically summarizes the distinction between additive and compensated Hall topology. The pair $(C,\mathcal{A})$ defined in Eqs.~\eqref{eq:C} and \eqref{eq:A} separates four regimes. A trivial compensated state has $(0,0)$. A hidden Hall altermagnet has $C=0$ and $\mathcal{A}\neq0$; local Hall-active masses exist, but ordinary transport cancels them. A partially compensated anomalous Hall regime has a finite, non-quantized response, usually because residual carriers, unequal sector weights or disorder spoil cancellation without creating a Chern insulator. An additive altermagnetic QAHE has non-zero integer $C$, a clean gap and chiral edge transport. Only this last regime should be called a quantized altermagnetic Chern insulator.

\section{Worked schematic example: two compensated Dirac sectors}
Consider two massive Dirac sectors with equal gap magnitude and opposite compensation parity, $\eta_1=+1$ and $\eta_2=-1$. Let their chiralities be equal, $\chi_1=\chi_2=+1$. This can occur when two Hall-active sectors contribute with the same sign for the same mass sign.

If altermagnetic exchange gives opposite projected masses, $\Delta_1=+m$ and $\Delta_2=-m$, then
\begin{equation}
C=\frac{1}{2}[+1-1]=0,
\qquad
\mathcal{A}=\frac{1}{2}[+1+1]=1.
\end{equation}
This is a hidden Hall state. Each sector is locally Hall active, yet the net Hall conductance vanishes. A surface-selective gate, valley-selective strain or asymmetric interface can reveal the hidden order by changing sector weights.

If a control parameter reverses the second mass so that $\Delta_1=\Delta_2=+m$, then
\begin{equation}
C=\frac{1}{2}[+1+1]=1,
\qquad
\mathcal{A}=\frac{1}{2}[+1-1]=0.
\end{equation}
The same compensated magnetic source has become additive in the Chern channel. The example is minimal, but it gives a direct experimental logic: tune the relative mass signs and follow the transition from hidden Hall compensation to quantized transport. If no sector gap opens, or if the Fermi level remains metallic, neither equation supports a QAHE claim.

\section{Materials routes and realistic failure modes}
Interfacial heterostructures provide the most controlled near-term platform. A topological insulator or topological crystalline insulator supplies inverted states, while an altermagnet supplies compensated exchange. The interface decides the projected mass. The layer Hall and crystal-symmetry routes show how compensated exchange can produce sector-resolved Hall channels before a full CI is reached.\cite{ref32,ref33,ref34} Recent studies of CrSb/(Bi,Sb)$_2$Te$_3$, altermagnetic proximity and altermagnetism-induced surface Chern states show why exchange projection is important in the study of quantum materials.\cite{ref35,ref36,ref37} These systems establish exchange-mass engineering, but quantized altermagnetic QAHE plateau is yet to be realized.

In this regard, intrinsic compounds impose a stricter materials constraint. A single phase must combine compensated altermagnetic order, strong SOC, band inversion, suitable filling and low disorder. Correlated layered altermagnets define microscopic spin--orbital platforms.\cite{ref38} Lieb-lattice models and metallic altermagnetic transitions sharpen the model-material connection, but they do not by themselves establish a CI.\cite{ref39,ref40,ref41}

Topological crystalline insulators are a natural setting, because valleys provide symmetry-resolved sectors. Monolayer IV--VI and related crystalline platforms show how mirror or valley degrees of freedom can host inverted bands.\cite{ref42,ref43} If altermagnetic exchange opens valley-selective masses, orientation and termination decide whether valley Hall responses add or cancel. Crystallography becomes a mass-sign control parameter, while sample preparation fixes the sector weights. Large spin splitting is insufficient if trivial bands cross $E_{\rm F}$ or if the mass pattern cancels in transport. Two-dimensional and van der Waals systems offer stronger control through gates, strain, sliding and twist.\cite{ref44,ref45,ref46} Floquet and electrically switchable routes further test whether the mass sign can be controlled dynamically.\cite{ref47,ref48} Ferroelectric, antiferroelectric and ferroelastic switching provide additional order-parameter handles for reversing sector masses.\cite{ref49,ref50,ref51}

The principal false-positive channels are well defined. The magnetic order may not be genuinely compensated. The spin splitting may not track the intended N\'eel vector. The observed gap may arise from hybridization, confinement or disorder rather than exchange. A metallic anomalous Hall response may be mislabeled as QAHE. Hidden Hall compensation may also be mistaken for triviality. A decisive experiment must exclude these channels rather than rely on a Hall loop alone.

\begin{figure*}[t]
\centering
\includegraphics[width=0.98\textwidth]{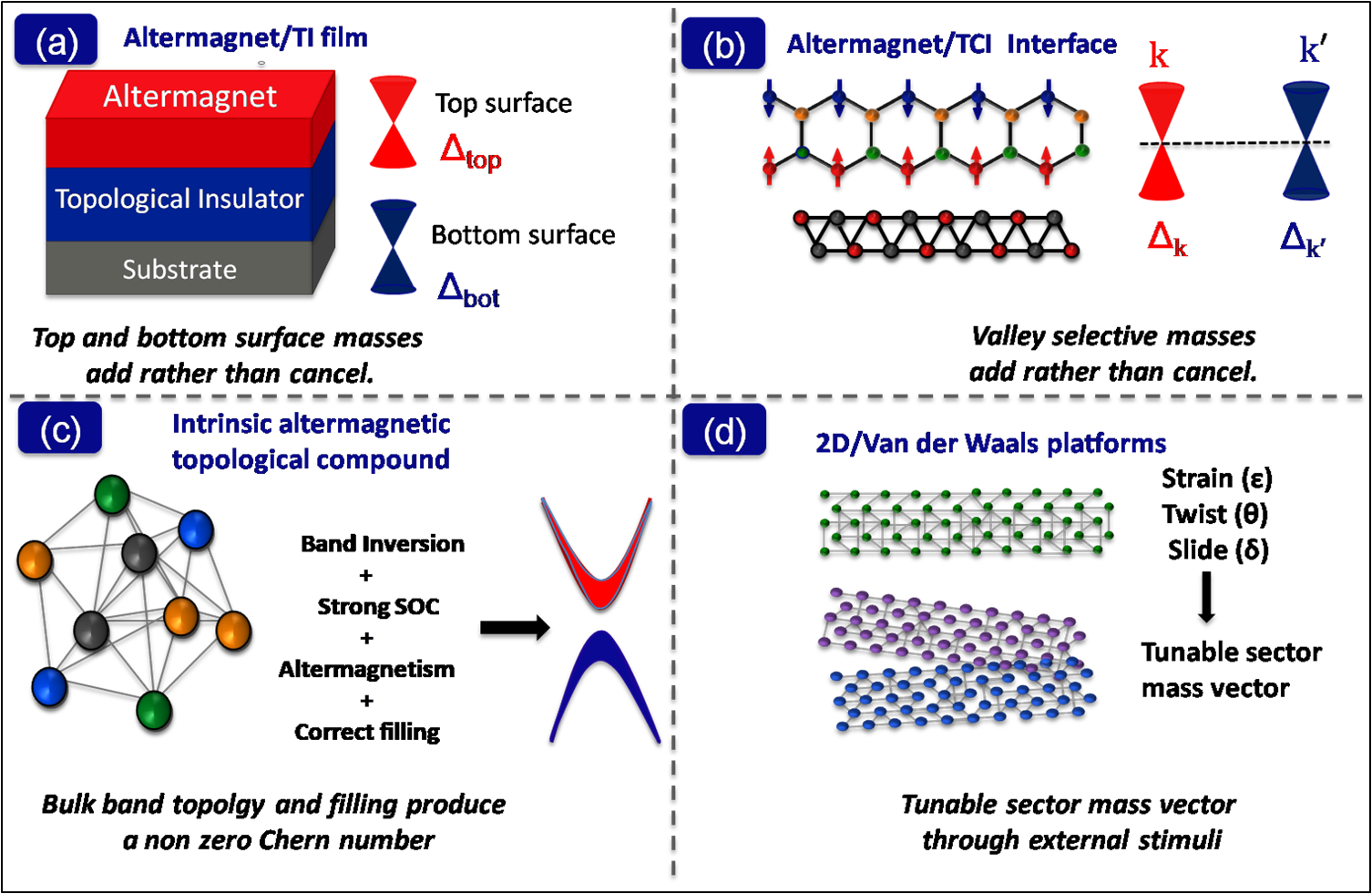}
\caption{\textbf{Materials roadmap for additive altermagnetic Chern matter.}
This figure summarizes the candidate routes comprising altermagnet/topological-insulator films, altermagnet/topological-crystalline-insulator interfaces, intrinsic altermagnetic topological compounds and two-dimensional platforms. Each route is a sectoral mass-vector problem. TI films measure surface masses on both top and bottom sides. The valley masses are tested in the interfaces of TCI. Intrinsic compounds test band inversion, SOC and filling. Two-dimensional systems perform tests for the sector weight of gates, strains, twists or slides.}\label{fig:materials-validation}
\end{figure*}

\section*{Box 1 | Minimal submission checklist}
\noindent\textbf{Projected mass.} Show that the altermagnetic exchange operator, together with SOC, gaps the intended topological sector. Do not rely only on spin splitting.

\medskip
\noindent\textbf{Additivity.} Determine whether surface, valley, layer or orbital masses add or cancel after chirality is included.

\medskip
\noindent\textbf{Controls.} Use interface orientation, registry, strain, gating, capping, thickness or N\'eel-vector manipulation to change mass signs or weights.

\medskip
\noindent\textbf{Negative controls.} Compare with nonmagnetic, differently oriented, ungated or thickness-varied structures to rule out hybridization, disorder and ordinary proximity effects.

\medskip
\noindent\textbf{Transport standard.} Do not call a state QAHE unless $R_{xy}$ is quantized, $R_{xx}$ is suppressed, the bulk is insulating and edge conduction is consistent with chirality.

\section{Experimental validation and testable predictions}
The first requirement is independent confirmation of compensated magnetic order. The magnetic symmetry can change due to substrate strain, off-stoichiometry and domain structures, which requires special care in the case of thin films. The second requirement is altermagnetic spin splitting associated with the N\'eel vector, ideally established by momentum-resolved spectroscopy or a symmetry-resolved probe.

The central criterion is sector-resolved mass spectroscopy. Sector-resolved probes, including ARPES, STS, magneto-optics, thickness scaling and interface-control series, are required to isolate the exchange-induced mass. The relevant quantity is the mass projected onto the active low-energy subspace. The extracted gap should be free of finite-size hybridization, quantum confinement, disorder self-energy and chemical-potential drift effects.

The compensation test should be based on the Berry-curvature weight. The sectoral mass terms should be redistributed by gate field, strain, interface registry or layer thickness. These controls should shift the system between hidden, partially compensated and additive Hall regimes. An invariant Hall response would prevent the altermagnetic mass channel from being the primary channel.

Transport quantization is used as the terminal criterion. A metallic anomalous Hall conductivity only indicates Berry-curvature imbalance; it is not evidence for QAHE. The decisive signature is the approach to $R_{xy}=h/(Ce^2)$ at zero field or in a remanent compensated state, while $R_{xx}$ converges to zero. The structure of chiral modes must be reproduced by nonlocal edge transport. These criteria are used to differentiate altermagnetic Chern matter from normal magnetic anomalous-Hall phases.

\section{Descendant phases and device relevance}
The immediate extension of additive altermagnetic Chern matter is proximity-induced superconductivity. Topological superconducting edge modes can exist at the boundaries of an $s$-wave proximitized chiral edge channel. This requires the superconducting pairing and the Chern mass to project onto the same low-energy sector. Lower stray-field constraints than with ferromagnetic platforms may be possible due to the compensated magnetic background.\cite{ref52,ref53,ref54,ref55}

A second route lies in higher-order topology. The projected mass can switch from positive to negative and vice versa along the crystallographic edges. These mass domains can bind hinge or corner modes. The phase can be tuned by Floquet driving, strain and structural switching. It can then be switched between trivial, hidden and additive Chern phases,\cite{ref47,ref56,ref57,ref58} all of which require the same mass audit: the modes must emerge from verified exchange projection. 
\section{Outlook}
Compensated order and altermagnetic spin splitting are not sufficient conditions for Chern matter. The essential operation is the projection of the exchange field onto a Hall-active mass channel. If this projection vanishes, the crystal may satisfy altermagnetic symmetry while remaining inactive in the Chern-insulator sector. If the projection is finite but sector-compensated, the system realizes hidden Hall order. If the projection is additive and opens a global gap, the system realizes an altermagnetic quantum anomalous Hall insulator.

This framework is experimentally falsifiable. Theory must resolve the projected exchange mass, sector chirality, compensation parity, and symmetry-breaking fields. Spin splitting alone is not a sufficient descriptor. A total Chern number alone is also insufficient. The spectroscopic gap opening has to be linked with the magnetic order experimentally. In addition to this, the quantized edge transport and compensation tuning need to be realized in the same setting. The immediate target is to identify materials in which compensated exchange produces an additive Chern mass.

\section*{Acknowledgements}
The author acknowledges the Institute of Physics (IOP), Bhubaneswar, for academic support and research environment.

\section*{Author contributions}
G.P.M. developed the projected-mass criterion, the compensated-topology diagnostic and the sector-resolved Chern-mass framework.

\section*{Competing interests}
The author declares no competing interests.

\end{document}